\documentclass[]{proarticle}
\usepackage{multicol}
\usepackage{graphicx}
\usepackage{amssymb}
\usepackage{epsfig}

\def\bea{\begin{eqnarray}}
\def\eea{\end{eqnarray}}
\renewcommand{\d}{{\rm d}}

\newcommand{\dagOm}{\Omega^\dagger}
\newcommand{\dotOm}{\dot{\Omega}}

\begin{document}
\title{On some feature and application of the Faddeev formulation of gravity}
\author{V.M. Khatsymovsky}
\maketitle
\address{Budker Institute of Nuclear Physics of Siberian Branch Russian Academy of Sciences, Novosibirsk, 630090, Russia.}
\eads{khatsym@gmail.com}
\begin{abstract}
In the Faddeev formulation of gravity, the metric is regarded as composite field, bilinear of $d = 10$ 4-vector fields. A unique feature is that this formulation admits the discontinuous fields. On the discrete level, when spacetime is decomposed into the elementary 4-simplices, this means that the 4-simplices may not coincide on their common faces, that is, be independent.

We apply this to the particular problem of quantization of the surface regarded as that composed of virtually independent elementary pieces (2-simplices). We find the area spectrum being proportional to the Barbero-Immirzi parameter $\gamma$ in the Faddeev gravity and described as a sum of spectra of separate areas.

According to the known in the literature approach, we find that $\gamma$ exists ensuring Bekenstein-Hawking relation for the statistical black hole entropy for arbitrary $d$, in particular, $\gamma = 0.39...$ for genuine $d = 10$.
\end{abstract}


MSC classes: 83C27; 53C05

\begin{multicols}{2}
\section{Introduction}
Here we briefly describe Faddeev formulation of gravity \cite{Fad}. Let the metric be composed of $d = 10$ 4-vector fields,

\begin{equation}                                                            
g_{\lambda\mu} = f^A_\lambda f_{\mu A}.
\end{equation}

\noindent The Greek indices $\lambda, \mu$, \dots = 1, 2, 3, 4, Latin capitals $A, B$, \dots = 1, \dots, $d$. Simple example: locally, 4D Riemannian space can be considered as a hyper-surface in the 10D Euclidean space. If $f^A (x)$ were its coordinates, then we would have

\begin{equation}                                                            
f^A_\lambda = \partial f^A / \partial x^\lambda.
\end{equation}

\noindent But now $f^A_\lambda (x)$ are a priori independent variables.

Now, in addition to the Christoffel connection, we can write an alternative affine connection

\begin{equation}                                                            
\Omega_{\lambda, \mu\nu} = f^A_\lambda f_{\mu A, \nu}, ~~~ \Omega^\lambda_{\mu\nu} = g^{\lambda\rho} \Omega_{\rho, \mu\nu},
\end{equation}

\noindent with torsion

\begin{equation}                                                            
T_{\lambda, [\mu\nu]} = \Omega_{\lambda, \mu\nu} - \Omega_{\lambda, \nu\mu}, ~~~ T^\lambda_{[\mu\nu]} = g^{\lambda\rho} T_{\rho, [\mu\nu]}
\end{equation}

\noindent and curvature, whose final expression is simple

\begin{eqnarray}                                                            
& & \hspace{-9mm} S^\lambda_{\mu, \nu\rho} = \Omega^\lambda_{\mu\rho, \nu} - \Omega^\lambda_{\mu\nu, \rho} + \Omega^\lambda_{\sigma\nu} \Omega^\sigma_{\mu\rho} - \Omega^\lambda_{\sigma\rho} \Omega^\sigma_{\mu\nu} = \Pi^{AB} \nonumber \\
& & \hspace{-9mm} \cdot (f^\lambda_{A, \nu} f_{\mu B, \rho} - f^\lambda_{A, \rho} f_{\mu B, \nu}), ~~~ \Pi_{AB} = \delta_{AB} - f^\lambda_A f_{\lambda B}.
\end{eqnarray}

The action is

\begin{eqnarray}                                                            
& & \int {\cal L} \d^4 x = \int g^{\lambda \nu} g^{\mu \rho} S_{\lambda \mu, \nu \rho} \sqrt {g} \d^4 x \nonumber \\ & & = \int \Pi^{AB} (f^\lambda_{A, \lambda} f^\mu_{B, \mu} - f^\lambda_{A, \mu} f^\mu_{B, \lambda}) \sqrt {g} \d^4 x,
\end{eqnarray}

\noindent $\Pi_{AB} = \delta_{AB} - f^\lambda_A f_{\lambda B}$ is projector onto the directions orthogonal to the four 10-vectors $f^\lambda_A$ (the "vertical" directions). Varying w. r. t. $f^\lambda_A$ and projecting onto the vertical directions we find

\begin{eqnarray}                                                            
& & b^\mu{}_{\mu A} T^\nu_{[\lambda \nu]} + b^\mu{}_{\lambda A} T^\nu_{[\nu \mu]} + b^\mu{}_{\nu A} T^\nu_{[\mu \lambda]} = 0, \nonumber \\ & & ~~~ b^A_{\lambda \mu} = \Pi^{AB} f_{\lambda B, \mu}.
\end{eqnarray}

\noindent This gives

\begin{equation}                                                            
T^\lambda_{[\mu\nu]} = 0
\end{equation}

\noindent almost everywhere. Then $\Omega^\lambda_{\mu\nu} = \Gamma^\lambda_{\mu\nu}$, and the action is the Einstein's one.
\section{The piecewise constant fields}\label{sec2}
The action does not contain any of the squared derivatives $(f^A_{\lambda, \mu})^2$ and therefore it is finite on the discontinuous  $f^A_{\lambda}$ and $g_{\lambda \mu}$. Consider a piecewise flat (composed of the flat 4-simplices) manifold. It can approximate (in the sense of curvature averaged over regions) general smooth manifold. Allowing discontinuities of the (induced on the 3D faces) metric $g_{\lambda \mu}$ means that the 4-simplices may not coincide on their common faces, that is, be independent. Also we can approximate general smooth manifold by {\it independent hypercubes} (were these hypercubes coinciding on their 3D faces, the metric would be strongly restricted).

To evaluate the action for this system, we consider \cite{II} the neighborhood of a triangle $\sigma^2_0$ where the 4-simplices $\sigma^4_1, \dots, \sigma^4_i, \dots, \sigma^4_n$ and their 3D faces $\sigma^3_i = \sigma^4_i \cap \sigma^4_{i + 1}$ ($\sigma^3_n = \sigma^4_n \cap \sigma^4_1$) are meeting, and $f^A_\lambda$ are independent in these 4-simplices.

In $\int {\cal L} \d^4 x = \int (f^\lambda_{A, \lambda} f^\mu_{B, \mu} - f^\lambda_{A, \mu} f^\mu_{B, \lambda}) \Pi^{AB} \sqrt {g} \d^4 x$ the Eq. $(f^\lambda_{A, \lambda} f^\mu_{B, \mu} - f^\lambda_{A, \mu} f^\mu_{B, \lambda})$ is $\propto$ $\delta$-function with support on $\sigma^2_0$ and the Eq. $\Pi^{AB} \sqrt {g}$ is $\propto$ Heaviside $\theta$-function in the neighborhood of $\sigma^2_0$ with discontinuities on $\sigma^3_i$, $i$ = 1, \dots, $n$. The product of $\delta$ and $\theta$-functions is not well-defined. However, the action is unambiguous in the leading order over variation of $f^A_\lambda$ from 4-simplex to 4-simplex. Vice versa, the action obtained correctly reproduces the continuum one in the continuum limit, when this variation of $f^A_\lambda$ tends to zero.
\section{Connection representation of the Faddeev action}
Write down Cartan-Weyl action for SO(10),

\begin{eqnarray}                                                            
& & S = \int f^\lambda_A f^\mu_B R^{AB}_{\lambda \mu} \sqrt{g} \d^4 x, \nonumber \\ & & \hspace{-7mm} R^{AB}_{\lambda \mu} = \partial_\lambda \omega^{AB}_\mu - \partial_\mu \omega^{AB}_\lambda + ( \omega_\lambda \omega_\mu - \omega_\mu \omega_\lambda )^{AB}.
\end{eqnarray}

\noindent Excluding $\omega^{AB}_\lambda$ via Eqs. of motion yields the Einstein action \cite{I}. Performing this operation with the following condition fulfilled,

\begin{equation}\label{omega-ff}                                           
\omega_{\nu}^{AB} f^\lambda_A f^\mu_B = 0,
\end{equation}

\noindent we get the Faddeev action.

We can generalize the above connection representation by adding the $1 / \gamma$-term to the action so that

\begin{eqnarray}                                                           
& & S = \int \left ( f^\lambda_A f^\mu_B + \frac{\epsilon^{\lambda \mu \nu \rho}}{2 \gamma \sqrt{g}} f_{\nu A} f_{\rho B} \right ) R_{\lambda \mu}^{AB} (\omega ) \sqrt{g} \d^4 x \nonumber \\ & & + \int \Lambda^\nu_{[\lambda \mu]} \omega_{\nu}^{AB} f^\lambda_A f^\mu_B \sqrt{g} \d^4 x,
\end{eqnarray}

\noindent where $\Lambda^\nu_{[\lambda \mu]}$ are Lagrange multipliers at the local gauge violating condition. Excluding $\omega^{AB}_\lambda$ via Eqs. of motion yields the generalized Faddeev action,

\begin{eqnarray}                                                           
& & S = \int \Pi^{AB} \left [ (f^\lambda_{A, \lambda} f^\mu_{B, \mu} - f^\lambda_{A, \mu} f^\mu_{B, \lambda}) \sqrt {g} \right. \nonumber \\ & & \left. - \frac{1}{\gamma} \epsilon^{\lambda \mu \nu \rho} f_{\lambda A, \mu} f_{\mu B, \rho} \right ] \d^4 x.
\end{eqnarray}

\noindent Using vertical components of the Eqs. of motion we get the Einstein action. It is remarkable that parameter $\gamma$, the analog of the Barbero-Immirzi parameter for the Cartan-Weyl form of the Einstein general relativity, provides nonzero contribution to the Faddeev action (in contrast to the case of Einstein gravity).
\section{Discrete connection representation for the Faddeev action}
To get this representation, we take 1) discrete connection representation for the Einstein action (that is, for Regge calculus), now with SO(10) connection, + 2) discrete form of the local gauge violating condition (\ref{omega-ff}). The 1) is the sum over triangles $\sigma^2$,

\begin{equation}                                                           
S^{\rm discr} = \sum_{\sigma^2} A(\sigma^2 ) \arcsin \left [ \frac{v_{\sigma^2 AB }}{2 A(\sigma^2 )} R^{AB}_{\sigma^2}{} ( \Omega ) \right ].
\end{equation}

\noindent Here $v_{\sigma^2 AB }$ is (dual) bi-vector of the triangle $\sigma^2$, $A(\sigma^2 )$ is the module of $v_{\sigma^2 AB }$ (the area), $R^{AB}_{\sigma^2}{} ( \Omega )$ is holonomy of the discrete connection, the product of SO(10)-matrices $\Omega^{AB}_{\sigma^3}$ over tetrahedra $\sigma^3$ sharing $\sigma^2$. Excluding $\Omega$ via Eqs. of motion we get Einstein (Regge) action \cite{II}.

The form of 2) is straightforward,

\begin{equation}                                                           
\Omega_{\sigma^3 AB} v^{AB}_{\sigma^2}{} = 0.
\end{equation}

\noindent Here $\sigma^2$ is the face of the one of two 4-simplices $\sigma^4$ containing $\sigma^3$ and in which $v_{\sigma^2 AB }$ is defined.

\medskip

To summarize, the versions of the Faddeev formalism considered can be continuum or discrete. Besides that, these can include connection (be of the I order) or be purely in terms of the tetrad or edge vectors (be of the II order). We have the following correspondence between the versions of the Faddeev gravity,

\begin{equation}\label{diagr}
\begin{array}{ccc} \mbox{ii) continuum} & \rightarrow & \mbox{iii) discrete} \\ \mbox{I order} & & \mbox{\hspace{3mm} I order} \\ \uparrow & & \downarrow \\ \mbox{i) continuum} & \rightarrow & \mbox{iv) discrete} \\ \mbox{II order} & & \mbox{\hspace{4mm} II order} \end{array}
\end{equation}

\noindent Above we have constructed the discrete connection representation considering logical transitions $\mbox{i)} \to \mbox{ii)} \to \mbox{iii)}$ from genuine Faddeev formulation $\mbox{i)}$, and then we can exclude connections via Eqs. of motion, $\mbox{iii)} \to \mbox{iv)}$. Then it turns out that we can reproduce the well-defined (leading over variation of $f^A_\lambda$ from 4-simplex to 4-simplex) part of the Faddeev action obtained by direct evaluation on the piecewise flat manifold, $\mbox{i)} \to \mbox{iv)}$. That is, the diagram (\ref{diagr}) is commutative.
\section{Discrete connection action on hypercubes}
As mentioned in Section \ref{sec2}, possibility of discontinuous metrics allows to operate with the hypercube minisuperspace decomposition of spacetime instead of the more complex and inconvenient simplicial decomposition. Summation over hypercubes is equivalent to summation over vertices (sites). The action is

\begin{eqnarray}                                                           
& & S^{\rm discr} = \frac{1}{8 \pi G} \sum_{\rm sites} \sum_{\lambda, \mu} \frac{\sqrt{ (f^\lambda )^2 (f^\mu )^2 - (f^\lambda f^\mu )^2}}{2 \sqrt{-\det \| f^\lambda f^\mu \|}} \nonumber \\ & & \arcsin \left[ \frac{f^\lambda_A f^\mu_B - f^\mu_A f^\lambda_B}{2 \sqrt{ (f^\lambda )^2 (f^\mu )^2 - (f^\lambda f^\mu )^2}} R^{AB}_{\lambda\mu} (\Omega ) \right] \nonumber \\ & & + \frac{1}{8 \pi G \gamma} \sum_{\rm sites} \sum_{\lambda, \mu} \frac{\sqrt{ (\epsilon^{\lambda \mu \nu \rho} f_{\nu A} f_{\rho B} )^2}}{2 } \arcsin \left[ \right. \nonumber \\ & & \hspace{-7mm} \left. \frac{\epsilon^{\lambda \mu \nu \rho} f_{\nu A} f_{\rho B} }{2 \sqrt{ (\epsilon^{\lambda \mu \nu \rho} f_{\nu A} f_{\rho B}  )^2}} R^{AB}_{\lambda\mu} (\Omega ) \right] + \sum_{\rm sites} \sum_{\lambda, \mu, \nu} \Lambda^\lambda_{[\mu \nu]} \Omega^{AB}_\lambda \nonumber \\ & & \cdot (f^\mu_A f^\nu_B - f^\nu_A f^\mu_B)
\end{eqnarray}

\noindent \cite{III}, $R_{\lambda\mu} (\Omega ) = \dagOm_\lambda (T^{\rm T}_\lambda \dagOm_\mu) (T^{\rm T}_\mu \Omega_\lambda) \Omega_\mu$, $\lambda, \mu, \dots$ label coordinate (hypercube) directions, $T_\lambda$ is translation operator along the edge $x^\lambda$ to the neighboring site. Now we assume Minkowsky metric signature (+, +, +, --) and $\Omega_\lambda \in $ SO(d-1,1).
\section{Continuous time limit}
The manifold is viewed as constructed of a set of 3D leaves of similar (here cubic) structure labeled by a parameter $t$. The difference $\d t$ for the neighboring leaves is assumed infinitesimal, and different variables are assumed to have certain orders of magnitude in $\d t$,

\begin{eqnarray}                                                           
& & f^A_0 = O( \d t), ~~~ \Omega_0 = 1 + O( \d t), \nonumber \\ & & T_0 = 1 + \d t \frac{\d }{\d t} + O((\d t)^2).
\end{eqnarray}

\noindent In particular, we can find the kinetic term $p \dot{q}$ in the resulting Lagrangian \cite{III},

\begin{eqnarray}                                                           
& & \hspace{-7mm} S^{\rm discr} \! = \!\! \int \!\! \frac{\d t}{16 \pi G} \sum_{\rm sites} \sum_\lambda \left [ \sqrt{-\det \| g_{\lambda \mu} \|} (f^0_A f^\lambda_B - f^\lambda_A f^0_B) \right. \nonumber \\
& & \left.  + \frac{1}{\gamma} \epsilon^{0 \lambda \mu \nu} f_{\mu A} f_{\nu B} \right ] (\dagOm_\lambda \dotOm_\lambda)^{AB} + \dots .
\end{eqnarray}
\section{Area quantization}
The quantization of the surface area was first discussed in the continuum general relativity theory, namely, using Ashtekar variables as early as in the work \cite{Ash}. There expression for the operator of the surface area requires point splitting regularization, and in order to preserve gauge invariance, this expression should be modified by introducing the path ordered exponents of the connection field operator. Then evaluation of the area operator on certain set of states in the Hilbert space of states (loop states) gives a discrete set of values.

The surface area operators in terms of the {\it discrete} Ashtekar type variables were considered in Ref. \cite{Loll}.

Now in the discrete framework, we simply need to quantize elementary area, or the area of the 2-simplex (triangle). Area bivectors are canonically conjugate to the orthogonal connection matrices. This fact leads to quantization of the elementary area in qualitative analogy with the quantization of angular momentum that is canonically conjugate to an orthogonal matrix of rotation in three-dimensional space. Now we are faced with the $(d-2)$-dimensional angular momentum whose spectrum is well-known (see, e.g., Ref. \cite{Vilen}). We find that the elementary area spectrum is proportional to the Barbero-Immirzi parameter $\gamma$ in the Faddeev gravity \cite{III},

\begin{equation}                                                           
A \equiv 8 \pi l^2_p \gamma a(j), ~~~ a(j) = \frac{1}{2} \sqrt{j (j + d - 4)}, ~~~ l^2_p = G,
\end{equation}

\noindent with multiplicity

\begin{equation}                                                           
g(j) = \frac{(j + d - 5)!(2j + d - 4)}{j!(d - 4)!}.
\end{equation}
\section{Requirements from black hole physics}
The spectrum of horizon area in the loop quantum gravity (different from the spectrum of the generic surface area in that theory) was first used to calculate the black hole entropy in Ref \cite{ABCK}. The requirement that it be $(4 l^2_p)^{-1} A_{bh}$ where $A_{bh}$ is the horizon area (Bekenstein-Hawking relation) means an Eq. for $\gamma$. In order to take into account the so-called holographic bound principle for the entropy of any spherical nonrotating system including black hole \cite{Bek2,Tho,Sus}, in Refs \cite{Khr1,Glo,Khr2,Cor} the formula for the spectrum of the horizon area was chosen to coincide with the general formula for the spectrum of the surface area, and corresponding value of $\gamma$ found.

Now the elementary area spectrum obtained can be used to evaluate statistical black hole entropy \cite{Khr2}. This gives Eq. for $\gamma$ for the different $d$'s,

\begin{equation}                                                           
\sum_j g(j) e^{-2 \pi \gamma a(j)} = 1.
\end{equation}

\noindent Calculation for, e. g., genuine $d = 10$ gives

\begin{equation}                                                           
\gamma = 0.393487933....
\end{equation}

\noindent If one considers {\it global} embedding into external space, it may require as much as $d = 230$ dimensions \cite{230} and then

\begin{equation}
\gamma = 0.359772297....
\end{equation}

\noindent The dependence on $d$ is rather weak.
\section{Conclusion}
Faddeev formulation allows for discontinuous metrics. Thus, spacetime can be virtually composed of independent microblocks.

It can be recast to the connection representation both in the continuum and discrete (piecewise flat minisuperspace) case.

Physically reasonable area spectrum is possible which is consistent with black hole physics.
\section*{Acknowledgement}
The author is grateful to Ya.V. Bazaikin, I. B. Khriplovich and I.A. Taimanov for valuable discussions. This research has been supported by the Ministry of Education and Science of the Russian Federation, Russian Foundation for Basic Research through Grant No. 11-02-00792-a and Grant 14.740.11.0082 of federal program "personnel of innovational Russia".

\end{multicols}

\begin{thebibliography}{99}
\bibitem{Fad}
 Faddeev L. D. 2011 {\it Theor. Math. Phys.} {\bf 166} 279-290.
\bibitem{II}
 Khatsymovsky V. M. 2012 {\it Faddeev formulation of gravity in discrete form} [arXiv:1201.0808 [gr-qc]].
\bibitem{I}
 Khatsymovsky V. M. 2012 {\it First order representation of the Faddeev formulation of gravity} [arXiv:1201.0806 [gr-qc]].
\bibitem{III}
 Khatsymovsky V. M. 2012 {\it On area spectrum in the Faddeev gravity} [arXiv:1206.5509 [gr-qc]].
\bibitem{Ash}
 Ashtekar A., Rovelli C. and Smolin L. 1992 {\it Phys. Rev. Lett.} {\bf 69} 237-240 [arXiv:hep-th/9203079].
\bibitem{Loll}
 Loll R. 1997 {\it Class. Quant. Grav.} {\bf 14} 1725-1741 [arXiv:gr-qc/9612068].
\bibitem{Vilen}
 Vilenkin N. Ya. 1968 {\it Special Functions and the Theory of Group
 Representations} Translations of Mathematical Monographs (Amer.
 Math.  Soc., Providence, Rhode Island) {\bf  22}.
\bibitem{ABCK}
 Ashtekar A., Baez J., Corichi A. and Krasnov K. 1998 {\it Phys.Rev.Lett.} {\bf 80} 904-907 [arXiv:gr-qc/9710007].
\bibitem{Bek2}
 Bekenstein J. D. 1981 {\it Phys. Rev. D} {\bf 23} 287-298.
\bibitem{Tho}
 't Hooft G. 1993 Dimensional Reduction in Quantum Gravity, in {\it Salam Festschrift} (Singapore) [arXiv:gr-qc/9310026].
\bibitem{Sus}
 Susskind L. 1995 {\it J. Math. Phys.} {\bf 36} 6377-6396 [arXiv:hep-th/9409089].
\bibitem{Khr1}
 Khriplovich I. B. and Korkin R. V. 2002 {\it J. Exp. Theor. Phys.} {\bf 95} 1-4; 2002 {\it Zh. Eksp. Teor. Fiz.} {\bf 95} 5-9 [arXiv:gr-qc/0112074].
\bibitem{Glo}
 Ghosh A. and Mitra P. 2005 {\it Phys. Lett. B} {\bf 616} 114-117 [arXiv:gr-qc/0411035].
\bibitem{Khr2}
 Khriplovich I.B. 2008 {\it Phys. Atom. Nucl.} {\bf 71} 671-680 [arXiv:gr-qc/0506082].
\bibitem{Cor}
 Corichi A., Diaz-Polo J. and Fernandez-Borja E. 2007 {\it Class. Quant. Grav.} {\bf 24} 243-251 [arXiv:gr-qc/0605014].
\bibitem{230}
 Nash J.F. 1956 {\it Ann. Math.} {\bf 63} 20-63.
\end{thebibliography}
\end{document}